\documentclass{article}
\usepackage[utf8]{inputenc}
\usepackage[margin=2.5cm]{geometry}
\usepackage{amsmath}
\usepackage{cite}
\usepackage{authblk}
\usepackage{graphicx}
\usepackage{xcolor}
\definecolor{link}{rgb}{0.0, 0.0, 0.0 }
\usepackage[bookmarks = true,
			pdfstartview = FitH,
			colorlinks = true,
			urlcolor=link,
			citecolor=link,
			linkcolor=link,
			hyperfootnotes=false]{hyperref}

\begin{document}

\title{Performing quantum entangled biphoton spectroscopy using classical light pulses}
\author[1,2]{Liwen Ko\thanks{liwen.jko@berkeley.edu}}
\author[1,2]{Robert L. Cook\thanks{rlcook@berkeley.edu}}
\author[1,2]{K. Birgitta Whaley\thanks{whaley@berkeley.edu}}
\affil[1]{Department of Chemistry, University of California, Berkeley, CA 94720, USA}
\affil[2]{Kavli Energy Nanoscience Institute at Berkeley, Berkeley, CA 94720, USA}
\maketitle

\begin{abstract}
    We show that for a class of quantum light spectroscopy (QLS) experiments using $n=0,1,2,\cdots$ classical light pulses and an entangled photon pair (a biphoton state) where one photon acts as a reference without interacting with the matter sample, identical signals can be obtained by replacing the biphotons with classical-like coherent states of light, where these are defined explicitly in terms of the parameters of the biphoton states. An input-output formulation of quantum nonlinear spectroscopy is used to prove this equivalence. We demonstrate the equivalence numerically by comparing a classical pump - quantum probe experiment with the corresponding classical pump - classical probe experiment. This analysis shows that understanding the equivalence between entangled biphoton probes and carefully designed classical-like coherent state probes leads to quantum-inspired classical experiments and provides insights for future design of QLS experiments that could provide a true quantum advantage.
\end{abstract}

\section{Introduction} \label{sec:introduction}
\par Spectroscopy using quantum light, in particular, using non-classical pulses containing individual or entangled pairs of photons, has attracted much interest in recent years, both theoretically and experimentally, due to its potential to exploit the non-classical properties of light to outperform classical spectroscopy \cite{Mukamel_Rev_Mod_Phys,Schlawin_2017_tutorial,Mukamel_2020_Roadmap,Szoke_2020_Review, Raymer_2021_tutorial,Scarcelli_2003,Yabushita_2004,Kalachev_2007,Schlawin_2012_double_excitation,Dorfman_2014_stimulated_Raman,Kalashnikov_2014_Plasmonic,Schlawin_2016,Kalashnikov_2016_infrared,Kalashnikov_2017_HOM,Ye_2021_chiral,Zhang_Scully_2022}. Quantum light spectroscopy (QLS) has been proposed to enable simplification of congested spectra \cite{Schlawin_2016}, to target specific doubly excited states \cite{Schlawin_2012_double_excitation}, to improve the signal-to-noise ratio of linear spectroscopy \cite{Kalashnikov_2014_Plasmonic}, and to measure dephasing rates with high temporal resolution \cite{Kalashnikov_2017_HOM}. Understanding the extent to which such QLS experiments provide a quantum advantage requires careful comparison with experiments using classical states of light. For example, the {relationship} between a quantum pump - quantum probe experiment and classical two-dimensional (2D) {spectroscopy experiments} is examined in \cite{Ishizaki_2020}.

\begin{figure}
    \centering
    \includegraphics[scale=0.7]{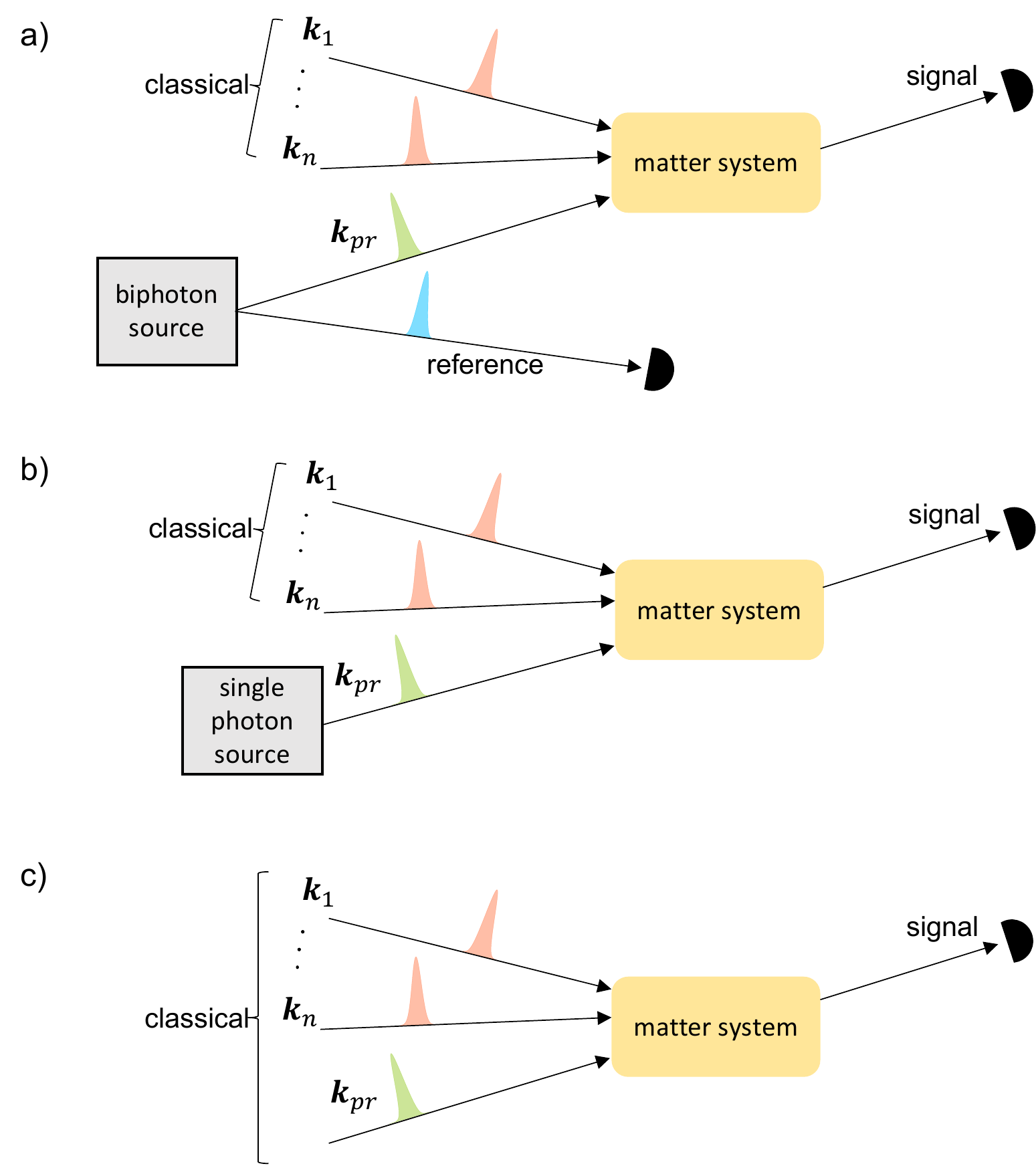}
    \caption{Spectroscopic schemes with $n=0,1,2,\cdots$ classical pulses and (a) a biphoton probe pulse with one of the photons acting as a reference without interacting with matter, (b) a single photon Fock state probe pulse, (c) a classical probe pulse containing one photon on average. The signal is measured in all cases in the direction of the probe field. The equivalence between schemes (a) and (b) is discussed in Sec. \ref{sec:biphoton_vs_single_photon}.
    Sec. \ref{sec:Fock_vs_classical} shows that when there is no phase matching of the $n$ classical pulses into the direction of the probe field, schemes (b) and (c) produce the same two-point correlation signal (e.g., photon flux, frequency-resolved photon count, or $g^{(1)}$ coherence function), providing equivalent measurable outcomes.
    }
    \label{fig:experimental_scheme}
\end{figure}

\par In this paper, we show that for a certain class of QLS experiments, the use of entangled photon pairs can be replaced with coherent states of light, which behave classically when normal-ordered field correlations are evaluated (see Fig. (\ref{fig:experimental_scheme})).  This is done in two steps. First, we show in Sec. \ref{sec:biphoton_vs_single_photon} that for QLS experiments using entangled photon pairs with one photon being measured without interacting with the matter system \cite{Dorfman_2014_stimulated_Raman,Scarcelli_2003,Kalachev_2007,Kalashnikov_2014_Plasmonic,Schlawin_2016,Yabushita_2004,Ye_2021_chiral,Zhang_Scully_2022}, the entangled photon pair can be replaced with a specially designed single photon Fock state, since measuring one photon effectively collapses the other photon into a single photon state. This has been pointed out previously by \cite{Stefanov_2017} in the context of metrology, and an analysis of the quantum information that one can extract from a two-level system using an entangled photon pair versus that extracted using a single photon Fock state has been provided in \cite{Francesco_arxiv}. Thus two-photon entanglement offers no true quantum advantage in these QLS experiments.
Nevertheless, there may still be practical advantages when using such entangled photon pairs with one photon acting as a reference without interacting with matter. For example, pure single photon Fock states are more difficult to produce experimentally than entangled photon pairs \cite{producing_pure_single_photon}, and a visible idler photon may be easier to detect than an infrared signal photon \cite{Kalashnikov_2016_infrared}.
\par 
Second, and this is the main theoretical result of the paper, we show in Sec. \ref{sec:Fock_vs_classical} that for a class of experiments using $n$ classical pulses ($n=0,1,2,\cdots$) and a single photon Fock state probe pulse, identical signals can also be obtained using a coherent state pulse having the same temporal profile and containing one photon on average. 
%Furthermore, the signal can be amplified by a factor equal to the average number of photons if one uses a coherent state probe containing many photons on average, making a coherent state probe far more preferable for experimental implementation.
%\bwcom{How about the following rewriting of the preceding sentence... somewhat expanded into 3 sentences as well for greater emphasis...}
Furthermore, if one uses a coherent state probe with the same temporal profile but containing many photons on average, the signal can be amplified by a factor equal to the average number of photons. Taken together with the equivalence of biphoton and single Fock state probes, this means that the spectral features obtained from experiments with biphoton probe pulses can be exactly reproduced and also amplified by carefully designed coherent state probe pulses. Such quantum-inspired coherent state probes are much simpler to implement and are thus far more preferable than biphoton states for experimental implementation.
\par 
We focus our analysis here on spectroscopy experiments for which the signal is measured in the direction of the probe.
For the case of a single classical pump, $n=1$, this allows direct comparison with conventional classical pump-probe spectroscopy and the entangled biphoton probe version of this that was proposed in \cite{Schlawin_2016}.
Other spectroscopy experiments where the signal is measured in directions other than the probe are not considered here, but can be analyzed similarly using the method we present here. For the equivalence between a single photon Fock state and a single photon coherent state to hold, we require that the classical pulses are incident from different directions than the probe pulse, and that there is no phase matching of the classical pulses into the direction of the probe pulse. In fact, the latter requirement includes the former as a special case. Neither of these are onerous requirements for experiments. 
\par 
In this work, we restrict the signal detection to normal-ordered two-point correlation measurements that contain one creation operator and one annihilation operator in the transmitted probe field, for example, photon flux $\langle a^\dagger(t)a(t)\rangle$, frequency-resolved photon count $\langle a^\dagger(\omega)a(\omega)\rangle$, or the unnormalized $g^{(1)}$ correlation function $\langle a^\dagger(t_2)a(t_1)\rangle$. We note that $g^{(1)}$ is complex-valued and therefore not a quantum mechanical observable in the strict sense, but the real and imaginary parts of $g^{(1)}$ can be measured separately using, for example, a Mach-Zehnder interferometer \cite{Loudon_2000_book}. The detection of higher-order coherence functions such as $g^{(2)}$ (a four-point correlation function) is not considered here.
The key reason for the equivalence between the Fock state probe and the coherent state probe in this class of experiments lies in the fact that they have the same two-point correlation function $\langle a^\dagger(t_2)a(t_1)\rangle$. While the one-point correlation functions $\langle a(t)\rangle$, $\langle a^\dagger(t)\rangle$, and other two-point correlation functions such as $\langle a(t_2)a(t_1)\rangle$ and $\langle a^\dagger(t_2)a^\dagger(t_1)\rangle$ are different for the two probes, their corresponding signals appear only in other phase matching conditions and do not appear in the direction of the probe field. So measuring the signal in this direction, as indicated in Fig.~(\ref{fig:experimental_scheme}), isolates signals that are dependent only on $\langle a^\dagger(t_2)a(t_1)\rangle$ and thereby ensures the desired equivalence. We note that this result is a generalization of our previous result \cite{Ko_2022} that the output photon flux is the same under single photon Fock state and single photon coherent state excitation of a matter system in the ground state. Combining the results in Sec. \ref{sec:biphoton_vs_single_photon} and \ref{sec:Fock_vs_classical}, we can then establish a class of QLS experiments that can be equivalently performed using only classical light.

\par In Sec. \ref{sec:numerical_example}, we use the classical pump - quantum probe experiment described in \cite{Schlawin_2016} as an example to numerically demonstrate how to perform the same experiment using only classical light to obtain the identical pump-probe spectrum. Concluding statements are presented in Sec. \ref{sec:conclusion}.

\section{Equivalence between signals from biphoton and single photon Fock state probes}
\label{sec:biphoton_vs_single_photon}
Consider an experiment where one probes a matter system using an entangled photon pair, whose density matrix is denoted as $\rho_{AB}$. Photon A 
{(e.g., the green pulse in Fig. (\ref{fig:experimental_scheme}a)) }
interacts with the system and the resulting output optical field is measured subsequently. Photon B {(e.g., the blue pulse in Fig. (\ref{fig:experimental_scheme}a))} is measured directly, without interacting with anything. 
Note that in this section focused on the equivalence between signals from biphoton and single photon Fock state probes, we shall impose no restriction on the observables to be measured in each of the two photon fields. 
For each realization of the experiment, a joint measurement of both photon fields is recorded as {$(\alpha,\beta)$}, where $\alpha$ represents the signal in photon field A in that experimental realization, {(e.g., whether or not a photon is present, or the measured frequency of the photon)}, and $\beta$ represents the measurement outcome of photon field B in the same experimental realization. Averaging over the signal $\alpha$ for fixed $\beta$ with repeated measurements, one obtains the final reference-averaged signal as $(S, \beta)$, where $S$ is the averaged signal of photon field A. 
Each $\beta$ corresponds to a value of the averaged signal $S$, so we shall henceforth abbreviate $(S, \beta)$ as $S_\beta$, representing the averaged signal of photon field A that is conditioned on the measurement outcome $\beta$ of photon field B. It is sometimes suggested that the correlation between the pair of photons A and B enhances the signal $S_\beta$ \cite{Schlawin_2016,Zhang_Scully_2022}. However, we show below that the conditional signal $S_\beta$ can in fact be constructed alternatively using just single photon states that are parameterized by $\beta$. In other words, in the experimental scheme of Fig. (\ref{fig:experimental_scheme}a), quantum entanglement between the two probe photons offers no fundamental advantage in learning about the matter system, since exactly the same results can be obtained using just single photon states. This has also been pointed out by Stefanov in \cite{Stefanov_2017}. 

\par To derive the single photon state that produces the same signal $S_\beta$, we first note that measuring photon B projects the photon pair state to $\Pi_{\beta}\rho_{AB}\Pi_{\beta}$, where $\Pi_{\beta}$ is the projector onto the eigenspace of the measurement outcome $\beta$. Since no further measurement is performed on photon B, photon A is then completely characterized by the reduced density matrix obtained by tracing the projected state over photon B: 
\begin{equation}
    \rho_{A|\beta}=\mathcal{N}\text{Tr}_{\text{B}} \big(\Pi_{\beta}\rho_{AB}\Pi_{\beta}\big).
    \label{Eq:single_photon_state_conditioned_j0}
\end{equation}
Here $\mathcal{N}=1/\text{Tr}(\Pi_{\beta}\rho_{AB}\Pi_{\beta})$ is a normalization factor to ensure unit trace. Eq.~(\ref{Eq:single_photon_state_conditioned_j0}) tells us that measuring the reference photon field B with outcome $\beta$ effectively collapses the input field of photon A into the single photon state $\rho_{A|\beta}$.
Therefore the conditional signal $S_\beta$ can also be obtained exactly by probing the system with the single photon state $\rho_{A|\beta}$.
\par
As an example, consider the frequency-entangled photon pair
\begin{equation}
    |\Psi\rangle = \int d\omega_A\int d\omega_B \Phi(\omega_A, \omega_B) a^\dagger_A(\omega_A)a^\dagger_B(\omega_B)|\text{vac}\rangle,
\label{Eq:biphoton_state}
\end{equation}
where $\Phi(\omega_A,\omega_B)$ is the biphoton wavefunction \cite{Loudon_2000_book},  $a_A(\omega_A)$ ($a_B(\omega_B)$) is the bosonic annihilation operator of frequency mode $\omega_A$ ($\omega_B$) in photon field A (B), and $|\text{vac}\rangle$ is the vacuum state of both fields. The operators $a_j(\omega)$ satisfy the bosonic commutation relations: $[a_j(\omega), a_{j'}(\omega')]=[a^\dagger_j(\omega), a^\dagger_{j'}(\omega')]=0$ and $[a_j(\omega), a^\dagger_{j'}(\omega')]=\delta_{j,j'}\delta(\omega-\omega')$. If we condition the experiment on measuring photon B at some reference frequency $\omega_B=\omega_r$, then the corresponding projection operator $\Pi_{\omega_r}$ is proportional to the outer product of the unnormalized state $a^\dagger_B(\omega_r)|\text{vac}\rangle$ and its adjoint, i.e.,
\begin{equation}
    \Pi_{\omega_r} \propto a^\dagger_B(\omega_r)|\text{vac}\rangle_B \langle \text{vac}|a_B(\omega_r),
\end{equation}
where $|\text{vac}\rangle_A$ or $|\text{vac}\rangle_B$ denotes the vacuum state for the photon field A or B. The projected photon pair state becomes 
\begin{equation}
    \Pi_{\omega_r}|\Psi\rangle \propto \int d\omega_A \Phi(\omega_A, \omega_r)a^\dagger_A(\omega_A) a^\dagger_B(\omega_r)|\text{vac}\rangle,
\end{equation}
which turns out to be a product state between the two photon fields A and B in this case. Therefore the reduced state of photon field A, $\text{Tr}_B ( \Pi_{\omega_r}|\Psi\rangle\langle \Psi|\Pi_{\omega_r})$, is a pure state, i.e.,
\begin{equation}
    \rho_{A|\omega_r} = |\psi\rangle_{\omega_r}\langle\psi|, 
\end{equation}
with
\begin{equation}
    |\psi\rangle_{\omega_r} = \mathcal{N}_{\omega_r} \int \, d\omega_A \Phi(\omega_A, \omega_r)a^\dagger_A(\omega_A)|\text{vac}\rangle_A,
\label{Eq:conditional_single_photon_example}
\end{equation}
where $\mathcal{N}_{\omega_r}=\sqrt{1/\int d\omega |\Phi(\omega, \omega_r)|^2}$ is the normalization factor.
Now the conditional signal  can be alternatively obtained using the single photon state of Eq. (\ref{Eq:conditional_single_photon_example}).
Note that the frequency profile of this single photon state is explicitly given by evaluating the biphoton wavefunction $\Phi(\omega_A, \omega_B)$ at $\omega_B = \omega_r$.
\par
The equivalence between signals from biphoton and single photon Fock state probes can be understood in a slightly different way by considering the photon correlation functions. For example, if one is interested in some property of the photon field A, represented by the quantum operator $O_A$, given that a photon with a frequency of $\omega_r$ is observed in the photon field B, one would typically need to evaluate the correlation function \cite{Schlawin_2016,Zhang_Scully_2022}
\begin{equation}
    \langle \Psi|a^\dagger_B(\omega_r)O_A a_B(\omega_r) |\Psi\rangle.
\label{Eq:conditional_single_photon_state}
\end{equation}
Since $a_B(\omega_r)|\Psi\rangle = \int d\omega_A\,\Phi(\omega_A,\omega_r)a^\dagger_A(\omega_A)|\text{vac}\rangle = \mathcal{N}_{\omega_r}^{-1} |\psi\rangle_{\omega_r}|\text{vac}\rangle_B$, Eq. (\ref{Eq:conditional_single_photon_state}) is equal to the expectation value
\begin{equation}
    \mathcal{N}_{\omega_r}^{-2}\,_{\omega_r}\langle \psi|O_A |\psi\rangle_{\omega_r}
\end{equation}
with respect to the reduced single photon state $|\psi\rangle_{\omega_r}$,
up to a normalization constant $\mathcal{N}_{\omega_r}$ that can be determined from the biphoton wavefunction $\Phi(\omega_A, \omega_B)$.

\section{Equivalence between signals from single photon Fock state and single photon coherent state probes}
\label{sec:Fock_vs_classical}

We now consider the class of experiments where $n$ classical pump pulses (with wavevectors $\mathbf{k}_1, \cdots, \mathbf{k}_n$) and a single photon Fock state or a single photon coherent state probe pulse (with wavevector $\mathbf{k}_{\text{pr}}$), treated quantum mechanically, interact with a matter system. These are illustrated in Fig. (\ref{fig:experimental_scheme}b) and Fig. (\ref{fig:experimental_scheme}c), for a single photon Fock state probe and a single photon coherent state probe, respectively. The classical pulses are incident at different directions from the quantum probe pulse, with the directions selected so that there is no phase matching of the classical pulses into the direction of the single photon probe. These conditions can be summarized as 
\begin{equation}
    \mathbf{k}_{\text{pr}} \text{ not proportional to } b_1\mathbf{k}_1 \pm \cdots \pm b_n\mathbf{k}_n
\label{Eq:phase_matching_constraint}
\end{equation}
where $b_i=0, 1, 2, \cdots$ can be any non-negative integer, up to a reasonable number of orders of interaction. The case of $n=0$ corresponds to the linear absorption of the single photon probe pulse; the case of $n=1$ corresponds to a classical pump - single photon probe experiment. We place no restriction on the relative time-ordering of the pulses. The signal is restricted to be normal-ordered two-point correlations that contain one creation operator and one annihilation operator in the probe field, e.g., photon flux $\langle a^\dagger_{\text{pr,out}}(t)a_{\text{pr,out}}(t)\rangle$, frequency-resolved photon count $\langle a^\dagger_{\text{pr,out}}(\omega)a_{\text{pr,out}}(\omega)\rangle$, or the $g^{(1)}$ coherence function $\langle a^\dagger_{\text{pr,out}}(t_2)a_{\text{pr,out}}(t_1)\rangle$. 
We claim that the final signal coming from the single photon Fock state probe
\begin{equation}
    |F_1\rangle = \int dt \,\xi(t)a^\dagger_{\text{pr}}(t)|\text{vac}\rangle
\label{Eq:single_photon_Fock_state}
\end{equation}
is equal to the signal from a coherent state probe
\begin{equation}
    |C_1\rangle = \exp \big(\int dt\, \xi(t)a^\dagger_{\text{pr}}(t) - \xi^*(t)a_{\text{pr}}(t)\big)|\text{vac}\rangle
\label{Eq:single_photon_coherent_state}
\end{equation}
having the same temporal profile $\xi(t)$ and containing on average a single photon. The temporal profile $\xi(t)$ is normalized according to $\int dt |\xi(t)|^2=1$. If the coherent state has $m$ photons on average, i.e., the state
\begin{equation}
    |C_m\rangle = \exp \big(\sqrt{m}\int dt \,\xi(t)a^\dagger_{\text{pr}}(t) - \xi^*(t)a_{\text{pr}}(t)\big)|\text{vac}\rangle,
\label{Eq:m_photon_coherent_state}
\end{equation}
then the probe field absorption and stimulated emission signal will be amplified by a factor of $m$. 

\par The key to this equivalence between {experiments} carried out with Fock state probes and coherent state probes is that pulses from these two probes have the same two-point correlation function $\langle a^\dagger_{\text{pr}}(t_2)a_{\text{pr}}(t_1)\rangle$. As already noted in Section~\ref{sec:introduction}, even though they have different two-point correlation functions $\langle a^\dagger_{\text{pr}}(t_2)a^\dagger_{\text{pr}}(t_1)\rangle$ and $\langle a_{\text{pr}}(t_2)a_{\text{pr}}(t_1)\rangle$ and different one-point correlation functions $\langle a^\dagger_{\text{pr}}(t)\rangle$ and $\langle a_{\text{pr}}(t)\rangle$, these other correlation functions do not contribute to the observed signal due to phase matching.
Together with the explicit parameterization of the coherent state pulse in terms of the single photon frequency profile $\xi(\omega) = \Phi(\omega, \omega_r)$ obtained from the biphoton state in Eq.~(\ref{Eq:conditional_single_photon_example}), this will allow replacement of a spectroscopic experiment using an entangled probe by experiments using a coherent state probe. 
\par 
We now prove the equivalence explicitly by analyzing the signals using an input-output formulation of quantum nonlinear spectroscopy. This approach is based on a perturbative expansion of the signal observables in the Heisenberg picture, distinct from the more common approach of perturbing the combined system plus field density matrix in the interaction picture \cite{Mukamel_Rev_Mod_Phys, Schlawin_2017_tutorial}. The input-output formulation simplifies the theoretical analysis by focusing on the signal observables and using standard results from  the input-output formalism of quantum optics \cite{Gardiner_Collett_1985,Combes_2017_review,Ko_2022}.
\par
Our analysis will focus on the frequency-resolved photon count signal $\langle a^\dagger_{\text{pr}}(\omega) a_{\text{pr}}(\omega)\rangle$. The analysis for other two-point correlation signals, such as photon flux and $g^{(1)}$ coherence function, follows almost identically. In the Heisenberg picture, the photon count of the transmitted probe at frequency $\omega$ is proportional to
\begin{equation}
   \int^\infty_{-\infty} dt_2 \int^\infty_{-\infty} dt_1 e^{i\omega(t_1-t_2)}\text{Tr}\big(\rho(-\infty) a^\dagger_{\text{pr, out}}(t_2)a_{\text{pr, out}}(t_1)\big).
\label{Eq:out_signal_Heisenberg}
\end{equation}
Here $\rho(-\infty)$ is the initial combined system plus probe field state, assumed to be a product state between the matter system $\rho_{M}$ and the field $\rho_{F}$, and the trace operator is evaluated over both the matter and the field degrees of freedom. $a_{\text{pr,out}}(t)$ is the output field operator of the probe field. This output field operator is the result of time-evolving the input field operator in the Heisenberg picture with the combined matter and field Hamiltonian, and thus it mixes the field and matter degrees of freedom \cite{Ko_2022}. The time domain field operator $a(t)$ is related to the frequency domain field operator $a(\omega)$ by the Fourier relation
\begin{equation}
    a(t) = \frac{1}{\sqrt{2\pi}}\int d\omega e^{-i\omega t} a(\omega).
\end{equation}
Therefore $a(t)$ also satisfies the bosonic commutation relations \cite{Loudon_2000_book}.
\par Although not necessary for the remaining derivation in this paper, we note that Eq. (\ref{Eq:out_signal_Heisenberg}) is expressed in \cite{Schlawin_2016} in a different form in the interaction picture as 
\begin{equation}
    \int dt_2 dt_1 e^{i\omega(t_1-t_2)}\text{Tr}\big(\rho(\infty) a^\dagger_{\text{pr}}(t_2)a_{\text{pr}}(t_1)\big),
    \label{Eq:out_signal_interaction}
\end{equation}
where $a_{\text{pr}}(t)$ is now the input field operator of the probe field. $\rho(\infty)$ is the combined system plus field state in the interaction picture, evolved to a time longer than $t_1$ and $t_2$.
{The term $\rho(\infty)$ is somewhat non-intuitive. To show the equality between Eqs. (\ref{Eq:out_signal_Heisenberg}) and (\ref{Eq:out_signal_interaction}), one considers how the input and output operators are related by unitary time-evolution operators. This is described in detail in Appendix \ref{app:Heisenberg_vs_interaction_signal}.} 
\par Under the dipole-electric field interaction and taking the zeroth order Hamiltonian as the pure system plus pure field Hamiltonian, we can write the interaction picture Hamiltonian as
\begin{align}
\begin{split}
    H(t) = & -ia_{\text{pr}}(t)L_{\text{pr}}^\dagger(t) + ia^\dagger_{\text{pr}}(t)L_{\text{pr}}(t) \\
    &+ \sum_{i=1}^{n} -i\alpha_i(t)L_i^\dagger(t) +i\alpha^*_i(t)L_i(t).
\label{Eq:Hamiltonian}
\end{split}
\end{align}
Eq.~(\ref{Eq:Hamiltonian}) consists of the system interaction with the quantum probe field, represented by the field operator $a_{\text{pr}}(t)$, and with $n$ other classical pulses, represented by their complex-valued coherent amplitudes $\alpha_i(t)$. The operators $L_{\text{pr}}$ and $L_i$ are the matter de-excitation components of the dipole operator corresponding to the probe field and the field of the i-th classical pulse, respectively. In the interaction picture, $L(t)=e^{iH_{\text{sys}}t} L e^{-iH_{\text{sys}}t}$ is a purely system operator (setting $\hbar=1$). 
\par Under the Hamiltonian of Eq. (\ref{Eq:Hamiltonian}), the input-output relation for the probe field is \cite{Gardiner_Collett_1985,Combes_2017_review,Ko_2022}
\begin{equation}
    a_{\text{pr,out}}(t) = a_{\text{pr}}(t) + L_{\text{pr,H}}(t),
\label{Eq:Input_output_relation}
\end{equation}
with $a_{\text{pr,out}}(t)$ the output probe field operator and $a_{\text{pr}}(t)$ the input probe field operator. Here $L_{\text{pr,H}}(t)$ is the Heisenberg evolved operator, defined as $U^\dagger(t)L_{\text{pr}}(t)U(t)$, where $U(t)$ is the time-evolution operator 
%$\mathcal{T}\exp -i\int^t_{-\infty} d\tau H(\tau)$ 
that solves the Schrodinger equation in the interaction picture, i.e., $dU(t)/dt = -iH(t)U(t)$. {The physical interpretation of Eq. (\ref{Eq:Input_output_relation}) is that the output electric field is equal to the input electric field plus the electric field generated by the matter dipole moment. } $L_{\text{pr}}(t)$, without the subscript ``H", will denote the operator in the interaction picture, which as noted above, is a purely system operator. In contrast, $L_{\text{pr,H}}(t)$ now mixes the system and field degrees of freedom. Performing a perturbative expansion on the backward Heisenberg equation of motion for $L_{\text{pr,H}}(t)$ \cite{unpublished}, we have
\begin{align}
\begin{split}
    L_{\text{pr,H}}(t) = & L_{\text{pr}}(t) -i\int^t_{-\infty}dt_1 [L_{\text{pr}}(t), H(t_1)] \\
    & + (-i)^2\int^t_{-\infty}dt_2\int^{t_2}_{-\infty}dt_1 [[L_{\text{pr}}(t), H(t_2)], H(t_1)] \\
    & + (-i)^3\int^t_{-\infty}dt_3\int^{t_3}_{-\infty}dt_2\int^{t_2}_{-\infty}dt_1 [[[L_{\text{pr}}(t), H(t_3)], H(t_2)], H(t_1)]\\
    & + \cdots.
\label{Eq:Heisenberg_perturbation}
\end{split}
\end{align}
The first term on right-hand side of 
Eq. (\ref{Eq:Heisenberg_perturbation}) can be interpreted as the matter dipole moment without interacting with the light, the second term as the matter dipole moment due to interacting with the field once, the third term as the matter dipole moment due to two interactions with the field, and so on. 
After expanding the commutators, each term becomes a product of a pure system operator and a pure field operator. Therefore the expectation values of $L_{\text{pr,H}}(t)$ with respect to an initial product state can be readily evaluated.

\par Substituting Eqs. (\ref{Eq:Input_output_relation}) and (\ref{Eq:Heisenberg_perturbation}) into the {$\text{Tr}$ operator in Eq. (\ref{Eq:out_signal_Heisenberg}), we obtain the following expansion:}
\begin{align}
\begin{split}
    \text{Tr}\big(\rho(-\infty) a^\dagger_{\text{pr,out}}(t_2)a_{\text{pr,out}}(t_1) \big)
    &= \Big\langle a^\dagger_{\text{pr,out}}(t_2)a_{\text{pr,out}}(t_1)\Big\rangle \\
    &= \bigg\langle \Big(a^\dagger_{\text{pr}}(t_2) + L^\dagger_{\text{pr}}(t_2) -i\int^{t_2}_{-\infty}d\tau [L^\dagger_{\text{pr}}(t_2), H(\tau)]+\cdots\Big)\\
    &\qquad\quad\Big(a_{\text{pr}}(t_1) + L_{\text{pr}}(t_1) -i\int^{t_1}_{-\infty}d\tau [L_{\text{pr}}(t_1), H(\tau)]+\cdots\Big)\bigg\rangle,
\label{Eq:signal_perturbative_expansion}
\end{split}
\end{align}
where we have adopted the conventional notation of using an angled bracket $\langle\hat{O}\rangle=\text{Tr}(\rho(-\infty) \hat{O})$ to denote the expectation value of an operator $\hat{O}$ with respect to the initial system plus field state $\rho(-\infty)$.
\par 
We now expand the right-hand side of Eq. (\ref{Eq:signal_perturbative_expansion}) in orders of $L_{\text{pr}}$. To show that $L_{\text{pr}}$ is indeed proportional to a small parameter, first notice that $L_{\text{pr}}$, $a_{\text{pr}}$, and the $L_{\text{i}}$ and $\alpha_i$ from the semi-classical terms of the Hamiltonian (Eq. (\ref{Eq:Hamiltonian})) all have the same dimension of $1/\sqrt{\text{time}}$ (setting $\hbar=1$).
Since $\langle L^\dagger_{\text{pr}}(t)L_{\text{pr}}(t)\rangle$ ($\langle L^\dagger_{i}(t)L_{i}(t)\rangle$) is at most equal to the spontaneous emission rate into the probe field ($i$\textsuperscript{th} classical field), where the maximum rate is obtained when the matter state is in the bright state of the corresponding field mode, we assign an order of magnitude value
\begin{equation}
    L \sim \sqrt{\eta/\tau_{\text{emission}}}
\end{equation}
to each $L$, where $\tau_{\text{emission}}$ is the time scale of spontaneous emission into the polarization of that field mode, and $\eta$ is the geometric factor of the field mode \cite{Ko_2022, Cook2022_arXiv}. $\eta$ is less than $1$ because a paraxial mode in an experiment usually covers only a small fraction of all light with the polarization of that paraxial mode. For a light pulse containing an average of $m$ photons, we have $\int dt \langle a^\dagger(t)a(t)\rangle = m$ (for classical pulses, we replace the operator $a(t)$ with the coherent amplitude $\alpha(t)$). 
Therefore we assign an order of magnitude value
\begin{equation}
    a(t),\alpha(t)\sim \sqrt{m/\tau_{\text{pulse}}}
\end{equation}
to each $a(t)$ or $\alpha(t)$, where $\tau_{\text{pulse}}$ characterizes the pulse duration.
In typical visible spectroscopy experiments with atomic and molecular samples, $\tau_{\text{emission}}\gg\tau_{\text{pulse}}$, so that the matter system dynamics is observable before spontaneous emission removes the excitation. Furthermore, since $m\gg 1$ for typical classical pulses and $m=1$ for the single photon probe pulse, we conclude that the magnitude of $L$ is much smaller than the magnitude of $a_{\text{pr}}$ and $\alpha_i$, justifying an expansion in powers of the $L$ operators. We then choose to expand Eq. (\ref{Eq:signal_perturbative_expansion}) only in orders of $L_{\text{pr}}$, since the orders of $L_i$ do not affect the main result, i.e., the equivalence of signals originating from a single photon Fock state probe and a single photon coherent state probe.
Furthermore, since $L_i$ always appears together with the classical pulse amplitude $\alpha_i$, the effect of $L_i$ is amplified by a factor of $\sqrt{m}$, so $L_{\text{pr}}$ becomes indeed the smallest parameter in the expansion of Eq. (\ref{Eq:signal_perturbative_expansion}). We now analyze the three lowest order contributions to the expansion.
\par
\textbf{Zeroth order term $(\sim{L_{\text{pr}}}^0)$.} The only zeroth order term in Eq. (\ref{Eq:signal_perturbative_expansion}) is $\langle a^\dagger_{\text{pr}}(t_2)a_{\text{pr}}(t_1)\rangle$, the transmitted probe without any interaction with matter. This expectation value is the same for both the single photon Fock state (Eq. (\ref{Eq:single_photon_Fock_state})) and the single photon coherent state (Eq. (\ref{Eq:single_photon_coherent_state})), namely
\begin{equation}
    \langle a^\dagger_{\text{pr}}(t_2)a_{\text{pr}}(t_1)\rangle = \xi^*(t_2)\xi(t_1),
\label{eq:zeroth_order}
\end{equation}
where $\xi(t)$ is the pulse shape.
For the $m$-photon coherent state (Eq. (\ref{Eq:m_photon_coherent_state})), Eq. (\ref{eq:zeroth_order}) is amplified by a factor of $m$.
\par \textbf{First order terms $(\sim{L_{\text{pr}}}^1)$.} Any first-order term in Eq. (\ref{Eq:signal_perturbative_expansion}) must be the expectation value of a product between $a^{(\dagger)}_{\text{pr}}$ and another term containing a single factor of $L_{\text{pr}}$, or its complex conjugate. In other words, only the semi-classical part of the Hamiltonian can contribute in the commutators of Eq. (\ref{Eq:signal_perturbative_expansion}); otherwise, there will be more than one factor of $L_{\text{pr}}$. Specifically, the first-order terms take the form 
\begin{align}
\begin{split}
\int d\tau_l\cdots d\tau_1 \bigg\langle &\colorbox{cyan}{$a^\dagger_{\text{pr}}(t_2)$}\\
&\Big[ \big[\cdots [\,\colorbox{yellow}{$L_{\text{pr}}(t_1)$}, \alpha_{i_l}^{(\pm)}(\tau_l)L^{(\mp)}_{i_l}(\tau_l)],\cdots \big], \alpha_{i_1}^{(\pm)}(\tau_1)L^{(\mp)}_{i_1}(\tau_1)\Big]\bigg\rangle,
\label{Eq:first_order_form}
\end{split}
\end{align}
and its complex conjugates. Here $l=0,1,2,\cdots$, and each $i$ index can denote any one of the $n$ classical field interactions. The probe field operators (i.e., $a_{\text{pr}}$ or $a^\dagger_{\text{pr}}$) are highlighted in blue, while the matter operator associated with the probe field (i.e., $L_{\text{pr}}$ or $L^\dagger_{\text{pr}}$) are highlighted in yellow. When $l=0$, Eq. (\ref{Eq:first_order_form}) reduces to $\langle a^\dagger_{\text{pr}}(t_2)L_{\text{pr}}(t_1)\rangle$. The notation $\alpha_{i}^{(\pm)}(\tau)L^{(\mp)}_{i}(\tau)$ means either $\alpha^*_i(\tau)L_i(\tau)$ or $\alpha_i(\tau)L^\dagger_i(\tau)$. Physically, Eq. (\ref{Eq:first_order_form}) represents the heterodyne measurement between the probe field (i.e., the $a^\dagger_{\text{pr}}$ in the first line) and the field generated by the matter polarization that is induced by the interactions with the classical fields (the second line). In all of the first order terms, the probe field expectation value factorizes out as $\langle a^\dagger_{\text{pr}}(t)\rangle$ or $\langle a_{\text{pr}}(t)\rangle$. These one-point correlation functions are zero for Fock state inputs and nonzero for coherent state inputs; therefore Eq. (\ref{Eq:first_order_form}) is different for Fock state and coherent state inputs. However, the optical signal generated by the matter polarization has the phase matching condition \cite{unpublished, Mukamel_book, hamm_zanni_book}
\begin{equation}
    \mathbf{k}_{\text{sig}} = \mathbf{k}_{i_1}\pm \cdots \pm \mathbf{k}_{i_l},
\label{Eq:frist_order_phase_matching}
\end{equation}
where the $k_i$ on the right-hand side are the wavevectors of the classical pulses. This means that $\mathbf{k}_{\text{sig}}$ must be in a different direction than the probe field direction $\mathbf{k}_{\text{pr}}$, due to our assumption of the beam geometry in Eq. (\ref{Eq:phase_matching_constraint}), i.e., the probe pulse is not phase matched with any of the classical pulses. Therefore the probe field signal of Eq. (\ref{Eq:first_order_form}) will vanish because it is not phase-matched to the matter polarization. At the molecular level, this means that in our beam geometry, the polarization from different molecules will generate destructively interfering signals and result in zero overall signal.
Hence the first order $(\sim{L_{\text{pr}}}^1)$ signal does not contribute to the probe field output.

\par \textbf{Second order terms $(\sim{L_{\text{pr}}}^2)$.} There are two types of second order terms. The first type is related to spontaneous emission and takes the form
\begin{align}
\begin{split}
    \text{Type 1:}\int d\tau_l \cdots d\tau_1 d\sigma_p\cdots d\sigma_1 \bigg\langle & \Big[\big[\cdots[\,\colorbox{yellow}{$L^\dagger_{\text{pr}}(t_2)$}, \alpha^{(\pm)}_{i_l}(\tau_l)L^{(\mp)}_{i_l}(\tau_l)],\cdots \big],\alpha^{(\pm)}_{i_1}(\tau_1)L^{(\mp)}_{i_1}(\tau_1)\Big] \\
    & \Big[\big[\cdots[\,\colorbox{yellow}{$L_{\text{pr}}(t_1)$}, \alpha^{(\pm)}_{j_p}(\sigma_l)L^{(\mp)}_{j_p}(\sigma_l)],\cdots \big],\alpha^{(\pm)}_{j_1}(\sigma_1)L^{(\mp)}_{j_1}(\sigma_1)\Big]\bigg\rangle.
\label{Eq:second_order_type1}
\end{split}
\end{align}
The integrand is a product of two nested commutators. Here $l$ and $p$ can take values of $0, 1, 2, \cdots$. Each of the $i$ and $j$ indices can be any one of the $n$ semi-classical interactions of the Hamiltonian. We take only the semi-classical part of the Hamiltonian in the commutators, since there is already one $L_{\text{pr}}$ in each of the two nested commutators. Otherwise Eq. (\ref{Eq:second_order_type1}) will contain more than two $L_{\text{pr}}$, becoming a higher-order term. In the case of $l=p=0$, Eq. (\ref{Eq:second_order_type1}) becomes $\langle L^\dagger_{\text{pr}}(t_2)L_{\text{pr}}(t_1)\rangle$: this represents spontaneous emission into the probe field without any interaction with the classical pulses. Since Eq. (\ref{Eq:second_order_type1}) contains no probe field operator (i.e., no $a_{\text{pr}}$ or $a^\dagger_{\text{pr}}$ terms), the expectation value is the same for all input field states, regardless of the phase matching conditions. For completeness, we note that the phase matching condition for these terms is \cite{unpublished}
\begin{equation}
    0 = \mathbf{k}_{i_1}\pm \cdots \pm \mathbf{k}_{i_l} \pm \mathbf{k}_{j_1}\pm \cdots \pm \mathbf{k}_{j_p}.
\end{equation}
\par The second type of second-order terms is related to absorption and stimulated emission, and has the form of
\begin{align}
\begin{split}
    \text{Type 2:}&\int d\tau_l\cdots d\tau_1 \bigg\langle \colorbox{cyan}{$a^\dagger_{\text{pr}}(t_2)$}\\
    & \quad\Big[\big[\cdots[[\cdots[\,\colorbox{yellow}{$L_{\text{pr}}(t_1)$}, \alpha^{(\pm)}_{i_l}(\tau_l)L^{(\mp)}_{i_l}(\tau_l)],\cdots ],\colorbox{cyan}{$a^{(\pm)}_{\text{pr}}(\tau_j)$}\colorbox{yellow}{$L^{(\mp)}_{\text{pr}}(\tau_j)$}\,],\cdots \big],\alpha^{(\pm)}_{i_1}(\tau_1)L^{(\mp)}_{i_1}(\tau_1)\Big] \bigg\rangle
\label{Eq:second_order_type2}
\end{split}
\end{align}
and its complex conjugate. In the nested commutator expression here, there is exactly one interaction with the quantum probe field. Physically, Eq. (\ref{Eq:second_order_type2}) represents the heterodyne measurement between the probe field (i.e., the $a^\dagger_{\text{pr}}$ in the first line) and the field generated by the matter polarization that is induced by one interaction with the quantum probe field and a number of interactions with the classical fields (the second line). The notation of the probe interaction term $a^{(\pm)}_{\text{pr}}(\tau_j)L^{(\mp)}_{\text{pr}}(\tau_j)$ stands for either the product $a^{\dagger}_{\text{pr}}(\tau_j)L_{\text{pr}}(\tau_j)$ or $a_{\text{pr}}(\tau_j)L^{\dagger}_{\text{pr}}(\tau_j)$. 
\par
When the probe field interaction is $a^\dagger_{\text{pr}}(\tau_j)L_{\text{pr}}(\tau_j)$, the probe field correlation in Eq. (\ref{Eq:second_order_type2}) factorizes out as $\langle a^\dagger_{\text{pr}}(t_2) a^\dagger_{\text{pr}}(\tau_j)\rangle$, 
which is zero for Fock state inputs and nonzero for coherent state inputs. Now the optical signal generated by the matter polarization has the phase matching condition of
\begin{equation}
    \mathbf{k}_{\text{sig}}= \mathbf{k}_{i_1}\pm \cdots - \mathbf{k}_{\text{pr}} \cdots \pm \mathbf{k}_{i_l},
\label{Eq:second_order_type1_phase_matching}
\end{equation}
where the right hand side contains only one probe field wavevector $k_{\text{pr}}$, and all other $k_i$ are the classical pulse wavevectors. But as discussed above, $\mathbf{k}_{\text{sig}}$ cannot be in the same direction as $\mathbf{k}_{\text{pr}}$ due to our assumption of the beam geometry in Eq. (\ref{Eq:phase_matching_constraint}). Therefore the final signal is not phase matched in the probe field direction $\mathbf{k}_{\text{pr}}$ and will vanish.
So neither a Fock state input nor a coherent state input will produce any signal from the $a^\dagger_{\text{pr}}(\tau_j)L_{\text{pr}}(\tau_j)$ interaction in this direction. 
\par On the other hand, when the probe field interaction in Eq. (\ref{Eq:second_order_type2}) is $a_{\text{pr}}(\tau_j)L^\dagger_{\text{pr}}(\tau_j)$, the field correlation now factorizes as $\langle a^\dagger_{\text{pr}}(t_2)a_{\text{pr}}(\tau_j)\rangle$, 
which is the same for both the single-photon Fock state and single-photon coherent state inputs, regardless of the phase-matching condition. These terms represent the transient absorption/stimulated emission of the probe field due to the interaction with the classical pulses.
In this case the phase matching condition of the optical signal generated by the matter polarization is now
\begin{equation}
    \mathbf{k}_{\text{sig}}= \mathbf{k}_{i_1}\pm \cdots + \mathbf{k}_{\text{pr}} \cdots \pm \mathbf{k}_{i_l},
\end{equation}
where the right hand side consists of only one probe field wavevector $k_{\text{pr}}$, and all other $k_i$ are the classical pulse wavevectors. We see that now if the classical pulse wavevectors cancel each other out pairwise, then we will have the correct phase matching condition of $k_{\text{sig}}=k_{\text{pr}}$ that results in a non-zero final signal in the probe field.
\par Due to the weak nature of the interaction between a single photon and a molecule (for example the probability for a chlorophyll molecule to absorb a single photon is at most on the order of $\sim 10^{-6}$ due to phonon dephasing \cite{Ko_2022, Herman_2018}), it is reasonable to truncate Eq. (\ref{Eq:signal_perturbative_expansion}) up to second order in $L_{\text{pr}}$. This second-order truncation corresponds to one interaction with the probe field in the language of classical nonlinear spectroscopy \cite{Mukamel_book}. 
\par
We may then combine the analysis for all of the terms up to second order in $L_{\text{pr}}$ (i.e., Eqs. (\ref{eq:zeroth_order}), (\ref{Eq:first_order_form}), and (\ref{Eq:second_order_type1}), and the two cases in Eq. (\ref{Eq:second_order_type2})). Doing this, we see first that while the first-order contribution Eq. (\ref{Eq:first_order_form}) and the first case of the second type of second-order contribution Eq. (\ref{Eq:second_order_type2}) yield different values for Fock state and coherent state inputs, neither of these terms appears in the final signal due to the phase matching constraint, so they cannot contribute to a difference between Fock and coherent state inputs.  In contrast, the zeroth-order contribution Eqs. (\ref{eq:zeroth_order}), the first type of the second-order contribution (\ref{Eq:second_order_type1}), and the second case of the second type of second order contribution Eq. (\ref{Eq:second_order_type2}) yield the same value for both Fock state and coherent state inputs, and these terms do have the correct phase matching condition to contribute to the final signal. Therefore, provided that the coherent state has the same temporal profile as the Fock state, a single photon Fock state probe and a single photon coherent state probe will produce exactly the same signal in the experimental setups of Fig. (\ref{fig:experimental_scheme}b) and (\ref{fig:experimental_scheme}c). Furthermore, a many photon coherent state probe with the same temporal profile will amplify the signals of Eq. (\ref{eq:zeroth_order}) and the second case of Eq. (\ref{Eq:second_order_type2}) by a factor of $m$, where $m$ is the average number of photons.

\section{Pump quantum-inspired probe (PQIP) spectroscopy}
\label{sec:numerical_example}
\begin{figure}
    \centering
    \includegraphics[scale=0.4]{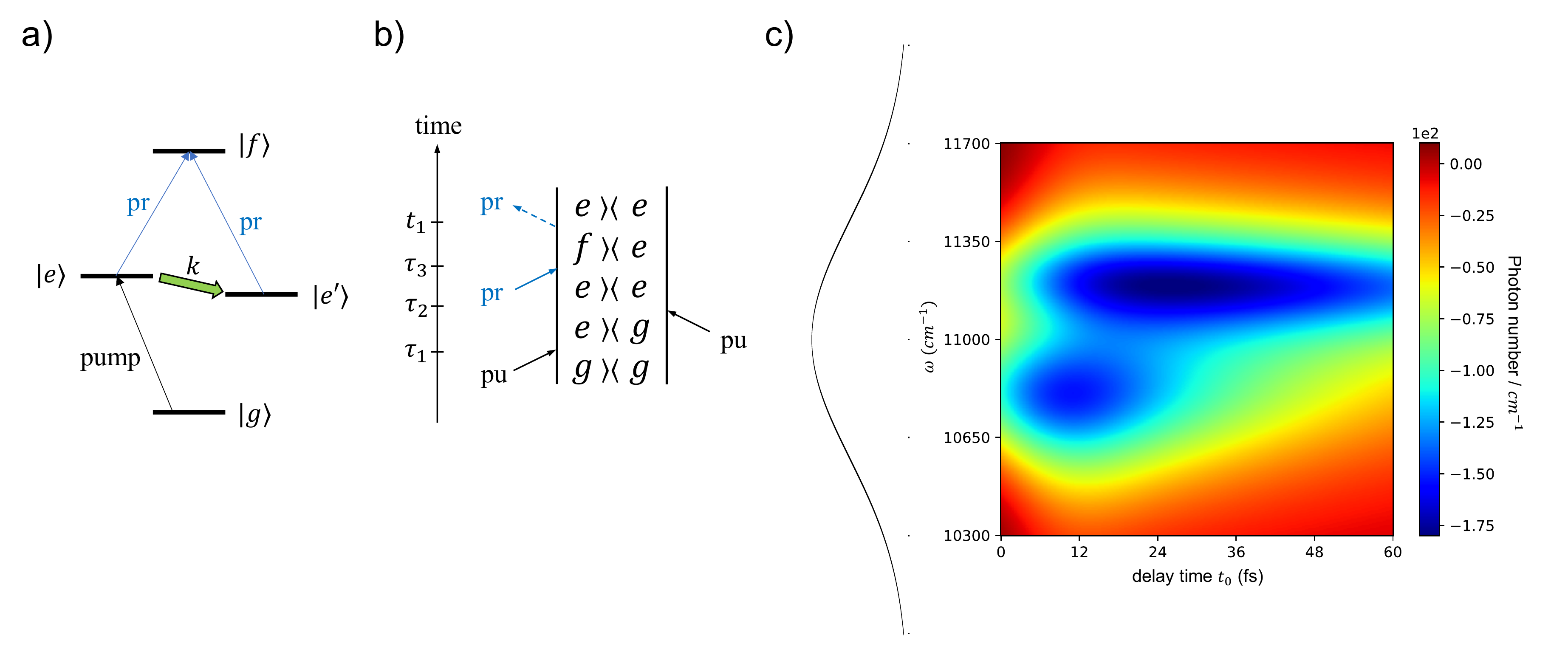}
    \caption{(a) Energy level scheme with four levels from \cite{Schlawin_2016}, which we use for our numerical example. The pump pulse is resonant to only the $|g\rangle \rightarrow |e\rangle$ transition. The probe pulse is resonant to only the $|e\rangle\rightarrow|f\rangle$ and $|e'\rangle\rightarrow|f\rangle$ transitions. (b) Double-sided Feynman diagram representing the excited state absorption of the pump-probe signal. The order of the first two pump interactions can be switched. (c) Transient absorption spectrum due to a conventional classical probe. The spectrum plots the change in the probe field frequency-resolved photon count $\langle a_{\text{pr}}^\dagger(\omega)a_{\text{pr}}(\omega)\rangle$ at frequency $\omega$, i.e., the signal photon number spectral density. This conventional classical probe has a Gaussian frequency profile $E(\omega)\propto e^{-(\omega-\omega_0)^2/2\sigma^2}$ ($\omega_0=11000\text{ cm}^{-1}$, $\sigma=600\text{ cm}^{-1}$) and contains on average $10^6$ photons. The frequency distribution $|E(\omega)|^2$ of the input probe pulse is plotted on the left of the spectrum.}
    \label{fig:classical_pump_probe}
\end{figure}

To demonstrate this equivalence between an entangled photon probe and a coherent state probe, we consider here the specific example of the classical pump - quantum probe experiment described theoretically in \cite{Schlawin_2016}, which corresponds to the case of a single classical pump pulse, i.e., $n=1$ in Fig. (\ref{fig:experimental_scheme}). We then compare this experiment to the corresponding classical pump - quantum-inspired classical probe experiment, which we shall refer to as ``pump quantum-inspired probe'' (PQIP). 
In this experiment, a delta-function classical pump first excites a four-level matter system from the ground state $|g\rangle$ to the singly excited state $|e\rangle$, which transfers the excitation to another lower-energy singly excited state $|e'\rangle$ irreversibly with a rate $k$ (see Fig. (\ref{fig:classical_pump_probe}a)). These energy transfer dynamics are monitored by the transient absorption of a probe pulse (delayed by time $t_0$ from the pump pulse) that excites $|e\rangle$ or $|e'\rangle$ into the doubly excited state $|f\rangle$. In~\cite{Schlawin_2016}, the probe pulse was taken to be either a classical pulse or an entangled photon pair. Fig. (\ref{fig:classical_pump_probe}c) shows the calculated transient absorption spectrum using a conventional classical probe pulse consisting of a single gaussian with frequency width $\sigma=600\text{ cm}^{-1}$, covering both transition frequencies from $|e\rangle$ and $|e'\rangle$ to $|f\rangle$. The structure of the two peaks centered at different delay times reveal the energy transfer dynamics from $|e\rangle$ to $|e'\rangle$.

\par In the case of a biphoton probe, the photon pair state $|\Psi\rangle$ is given by Eq. (\ref{Eq:biphoton_state}), with the biphoton wavefunction $\Phi(\omega_{\text{pr}}, \omega_r)$.
One photon (the reference photon) of the probe photon pair does not interact with the matter system and its frequency $\omega_r$ is measured. The other photon (the probe signal photon) interacts with the matter system and is frequency-resolved. For each $\omega_r$, there is a transient absorption spectrum as a function of the signal frequency $\omega$ and delay time $t_0$. As discussed in~\cite{Schlawin_2016}, due to the frequency correlation in the entangled photon pair, by selecting different values of $\omega_r$, one can target specific frequency windows of the transient absorption spectrum, thereby simplifying the spectrum. 
\par The theoretical analysis of these spectra obtained from biphoton pulses proceeds as follows. The pump-probe signal for a fixed reference photon frequency $\omega_r$ is the difference between the output and the input signals
\begin{equation}
    \langle a_r^\dagger(\omega_r)a_r(\omega_r) a_{\text{pr,out}}^\dagger(\omega)a_{\text{pr,out}}(\omega)\rangle - \langle a_r^\dagger(\omega_r)a_r(\omega_r) a_{\text{pr}}^\dagger(\omega)a_{\text{pr}}(\omega)\rangle.
\label{Eq:pump_probe_signal_reference_1}
\end{equation}
If no reference photon is used, the pump-probe signal becomes
\begin{equation}
    \langle a_{\text{pr,out}}^\dagger(\omega)a_{\text{pr,out}}(\omega)\rangle - \langle a_{\text{pr}}^\dagger(\omega)a_{\text{pr}}(\omega)\rangle.
\label{Eq:pump_probe_signal_no_reference_1}
\end{equation}
Applying Eqs. (\ref{Eq:Input_output_relation}) and (\ref{Eq:Heisenberg_perturbation}), the lowest order term of Eq. (\ref{Eq:pump_probe_signal_reference_1}), represented by the double-sided Feynman diagram of Fig. (\ref{fig:classical_pump_probe}b), is
\begin{align}
\begin{split}
&-\int^\infty_{-\infty} \frac{dt_2}{\sqrt{2\pi}} \int^\infty_{-\infty} \frac{dt_1}{\sqrt{2\pi}} e^{i\omega(t_1-t_2)} \int^{t_1}_{-\infty}d\tau_3 \int^{\tau_3}_{-\infty}d\tau_2 \int^{\tau_3}_{-\infty}d\tau_1 \\
&\qquad\qquad \Big\langle a_r^\dagger(\omega_r)a_r(\omega_r)a^\dagger_{\text{pr}}(t_2)a_{\text{pr}}(\tau_3)\Big\rangle\Big\langle L_{\text{pu}}(\tau_1) L_{\text{pr}}(t_1) L_{\text{pr}}^\dagger(\tau_3)  L_{\text{pu}}^\dagger(\tau_2)  \Big\rangle \alpha_{\text{pu}}^*(\tau_1)\alpha_{\text{pu}}(\tau_2) + \text{c.c.}
\label{Eq:pump_quantum_probe_signal}
\end{split}
\end{align}
Note that Eq. (\ref{Eq:pump_quantum_probe_signal}) originates from the second-order ($\sim L_{\text{pr}}^2$) expansion term of the form of Eq. (\ref{Eq:second_order_type2}). Substituting in the delta-function classical pump $\alpha_{\text{pu}}(t) \propto \delta(t)$, Eq. (\ref{Eq:pump_quantum_probe_signal}) is now proportional to
\begin{align}
\begin{split}
    -\int^\infty_{-\infty} d\omega' &\Big\langle a_r^\dagger(\omega_r)a_r(\omega_r)a^\dagger_{\text{pr}}(\omega)a_{\text{pr}}(\omega')\Big\rangle\\
    & \int^\infty_0 dt_1 \int^{t_1}_0 d\tau_3 \,e^{i\omega t_1}e^{-i\omega' \tau_3} \Big\langle L_{\text{pu}}(0) L_{\text{pr}}(t_1) L_{\text{pr}}^\dagger(\tau_3)  L_{\text{pu}}^\dagger(0)  \Big\rangle + \text{c.c.}
\label{Eq:pump_probe_signal_reference_2}
\end{split}
\end{align}
Since the field correlation function in Eq. (\ref{Eq:pump_probe_signal_reference_2}) with a time delay of $t_0$ evaluates to
\begin{equation}
    \big\langle a_r^\dagger(\omega_r)a_r(\omega_r)a^\dagger_{\text{pr}}(\omega)a_{\text{pr}}(\omega')\big\rangle=\Phi^*(\omega,\omega_r)\Phi(\omega', \omega_r)e^{i(\omega'-\omega)t_0},
\end{equation} 
The signal Eq. (\ref{Eq:pump_probe_signal_reference_2}) can be expressed compactly as \cite{Schlawin_2016}
\begin{equation}
    -\text{Re} \int d\omega' \Phi^*(\omega,\omega_r)\Phi(\omega', \omega_r)\Tilde{F}(\omega',\omega;t_0),
\label{Eq:pump_probe_simplified_signal}
\end{equation}
where 
\begin{equation}
    \Tilde{F}(\omega',\omega;t_0) = \int^\infty_0 dt_1 \int^{t_1}_0 d\tau_3 \,e^{i\omega (t_1-t_0)}e^{-i\omega' (\tau_3-t_0)} \Big\langle L_{\text{pu}}(0) L_{\text{pr}}(t_1) L_{\text{pr}}^\dagger(\tau_3)  L_{\text{pu}}^\dagger(0)  \Big\rangle
\end{equation}
is the frequency-domain matter correlation function defined in \cite{Schlawin_2016}.
\par
The detailed model of the matter system, the corresponding analytical form of $\Tilde{F}(\omega',\omega;t_0)$, and the analytical form of $\Phi(\omega_{\text{pr}},\omega_r)$ are discussed in \cite{Schlawin_2016} and summarized in Appendix \ref{app:numerical_parameter}. 
Similarly, if a single photon Fock state or a coherent state is used as the probe, then the pump-probe signal Eq. (\ref{Eq:pump_probe_signal_no_reference_1}) becomes \begin{equation}
    -\text{Re} \int d\omega' \xi^*(\omega)\xi(\omega')\Tilde{F}(\omega',\omega;t_0),
\label{Eq:pump_probe_simplified_signal_2}
\end{equation}
where $\xi(\omega)$ is the frequency profile of the probe pulse (see Eqs. (\ref{Eq:single_photon_Fock_state})-(\ref{Eq:m_photon_coherent_state})).
\par 
Comparing Eq. (\ref{Eq:pump_probe_simplified_signal}) and Eq. (\ref{Eq:pump_probe_simplified_signal_2}), we observe that if we choose a quantum-inspired coherent state probe with coherent amplitude $\xi_{\text{pr}}(\omega)=\Phi(\omega,\omega_r)$, the final signal is exactly proportional to Eq. (\ref{Eq:pump_probe_simplified_signal}) at a fixed reference photon frequency $\omega_r$.
Therefore the classical pump - quantum probe experiment can be exactly reproduced using a standard classical pump - classical probe setup, with the only additional feature of requiring a pulse shaper for the quantum-inspired classical probe pulse. The shape of the quantum-inspired classical probe is parameterized by $\omega_r$, together with the other parameters of the biphoton pulse (see Appendix \ref{app:numerical_parameter}).
\par
\begin{figure}
    \centering
    \includegraphics[scale=0.5]{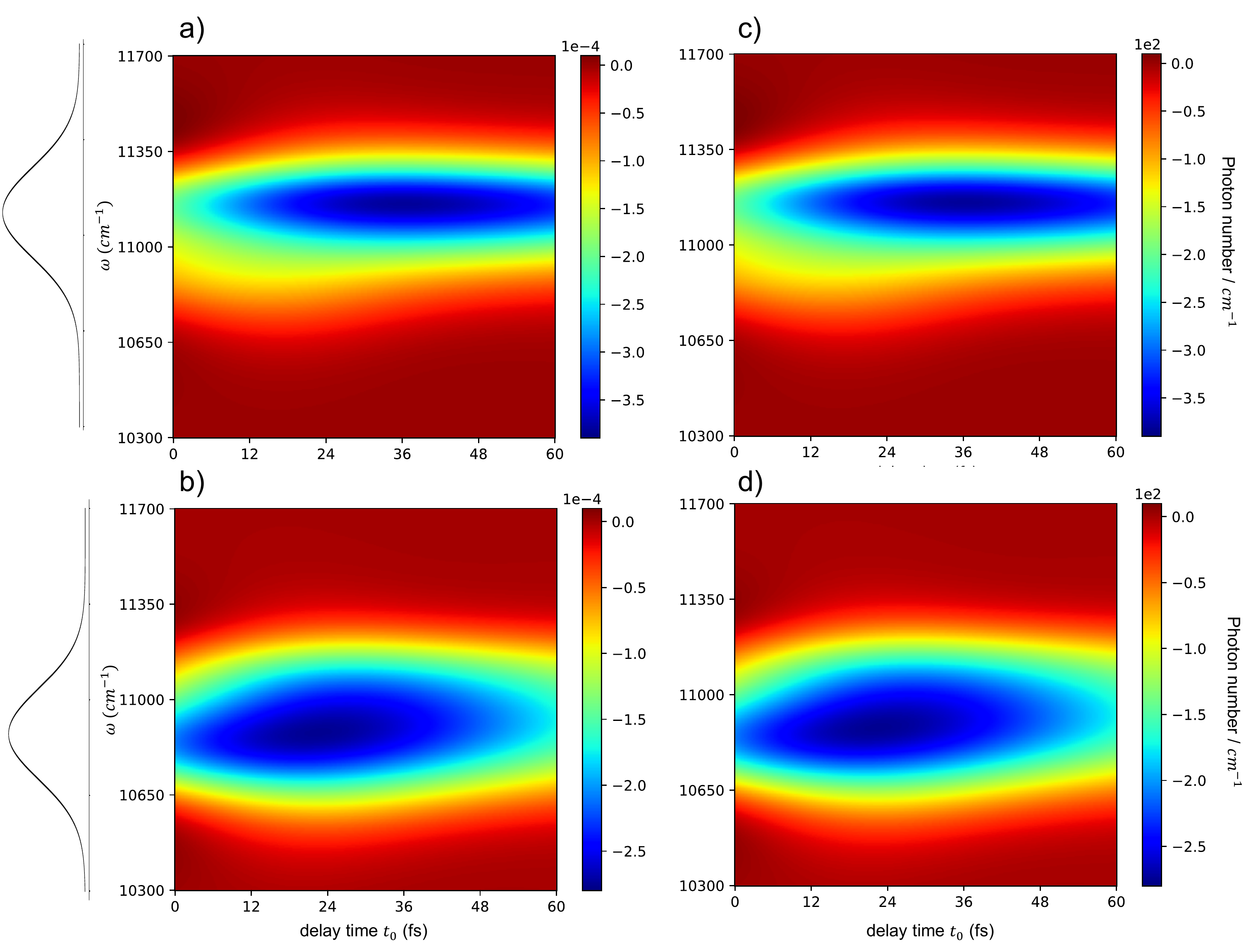}
    \caption{Transient absorption spectra obtained using (a-b) an entangled biphoton probe or (c-d) quantum-inspired classical probes. The signal is the change of the probe field frequency-resolved photon count $\langle a^\dagger(\omega)a(\omega)\rangle$ at frequency $\omega$, i.e., the signal photon number spectral density. In panels (a) and (b), the signal is conditioned on the reference photon frequencies of (a) $\omega_r=11400\text{ cm}^{-1}$ (b) $\omega_r=10400\text{ cm}^{-1}$. On the left of each spectrum is the frequency distribution $|\Phi(\omega, \omega_r)|^2$ of the probe single photon for fixed $\omega_r$.
    In panels (c) and (d), classical probes with frequency profiles $\xi(\omega)=\Phi(\omega,\omega_r)$ are used, where $\omega_r=11400\text{ cm}^{-1}$ in (c) and $\omega_r=10400\text{ cm}^{-1}$ in (d), corresponding to panels (a) and (b), respectively. The classical probe pulses contain $10^6$ photons on average, resulting in $10^6$ times signal amplification. Note that the scales of the color bars in (c) and (d) are $10^6$ larger than those in (a) and (b).}
    \label{fig:quantum_vs_classical_probe}
\end{figure}

The classical pump - quantum probe spectra with biphoton pulses, characterized by two choices of $\omega_r$, are shown in the left-hand panels (a) and (b) of Fig. (\ref{fig:quantum_vs_classical_probe}). The signal is detected in the probe beam direction, in accordance with the phase matching requirement discussed in Sec. \ref{sec:Fock_vs_classical}.
%\bwcom{BW: mention the phase matching condition here. 
The simplification of the spectra relative to the conventional pump-probe spectrum in Fig.~(\ref{fig:classical_pump_probe}) is immediately evident, with the two peaks now clearly resolved, permitting a more detailed analysis of the coupled dynamics underlying the two spectra.
\par
The corresponding PQIP spectra are shown in the right-hand panels (c) and (d) of Fig. (\ref{fig:quantum_vs_classical_probe}). In the numerical simulation, we use a classical probe containing an average of $m=10^6$ photons to  amplify the final signal by a factor of $10^6$. As noted above, this has the additional benefit of making the signal detection much easier experimentally than when using an entangled biphoton probe. When the amplified signals are normalized to the same reference value as that for panels (a) and (b), the left- and right-hand panels of Fig. (\ref{fig:quantum_vs_classical_probe}) are identical to within numerical precision, validating the PQIP analysis.
The specific quantum-inspired pulses that produce the same spectra as the biphoton probe with $\omega_r$ values in panels (a) and (b) of Fig.~(\ref{fig:classical_pump_probe}) are given explicitly in Appendix~\ref{app:numerical_parameter}.

\section{Conclusion}
\label{sec:conclusion}
We have shown that for a class of QLS experiments consisting of $n=0,1,2\cdots$ classical pulses and an entangled photon pair probe in the scheme of Fig. (\ref{fig:experimental_scheme}a), the use of the entangled photon pair can be replaced with a specially designed coherent state pulse, which behaves as classical light when normal-ordered field correlations are evaluated. The two main requirements for this equivalence to hold are the following: (1) there is no phase matching of the classical pulses into the direction of the probe field, and (2) signal measurement takes the form of (time-integrated) photon flux, frequency-resolved photon count, or $g^{(1)}(t)$ correlation function. 
\par
The class of experiments described in this paper is a subset of QLS experiments where the use of entangled photon pairs can be replaced with classical pulses. Whenever a biphoton input is used and a non-interacting reference photon $r$ is measured at frequency $\omega_r$, so that the signal consists of field correlation functions of the form 
$\langle a_r^\dagger(\omega_r)a_r(\omega_r)a^\dagger(t_2)a(t_1)\rangle$, we showed that the signal can be reproduced with coherent state pulses that are specifically designed for a given biphoton state and reference frequency.
In this work we also demonstrated the validity of the analysis by explicit calculations of the signal for a classical pump - entangled photon probe experiment, showing numerical equivalence with the signal obtained from a classical pump with a coherent state pulse that is constructed according to the formulation described in Sections 2-3 and Appendix B.
\par
Going beyond the scope of analysis in this paper, one may also consider the effect of photon noise  on spectroscopy with entangled photons \cite{Kalashnikov_2014_Plasmonic}. Here the signal-to-noise improvement offered by entangled photon pairs described in \cite{Kalashnikov_2014_Plasmonic}, can likely be achieved merely by using pulsed classical light in only the signal arm. If we go beyond the dipole-electric field interaction and allowing for Raman scattering interactions, one can also show that the intensity correlated Raman signal in \cite{Zhang_Scully_2022} and the (1,0) component of the interferometric stimulated Raman signal in \cite{Dorfman_2014_stimulated_Raman} can also be reproduced by classical pulses parameterized by the biphoton wavefunction and the reference photon frequency, since these signals depend on the same field correlation function as in Eq. (\ref{Eq:pump_quantum_probe_signal}). 
\par Finally, we note that while some QLS experiments can be reproduced using carefully designed classical light sources as shown here, at the same time the technologies for generation and detection of quantum light are maturing, raising the possibility of a new generation of QLS experiments.
The equivalence between entangled biphoton probes and classical-like coherent state probes shown in this work leads to a new category of quantum-inspired classical spectroscopy experiments, such as the pump quantum-inspired probe described in Sec. \ref{sec:numerical_example}. An understanding of the range of applicability of the equivalence demonstrated here will provide insights for future design of more powerful QLS experiments that cannot be replicated by suitably designed quantum-inspired classical pulses and that could provide a true quantum advantage for the study of electronic dynamics in complex systems. 

\section*{Acknowledgements}
L.K. was supported by the Kavli Energy NanoScience Institute (ENSI) Philomathia graduate fellowship. This project was supported by the Photosynthetic Systems program of U.S. Department of Energy, Office of Science, Basic Energy Sciences, within the Division of Chemical Sciences, Geosciences, and Biosciences, under Award No. DESC0019728.

\newpage
\appendix
\numberwithin{equation}{section}

\section{Relationship between Eqs. (\ref{Eq:out_signal_Heisenberg}) and (\ref{Eq:out_signal_interaction})}
\label{app:Heisenberg_vs_interaction_signal}
To show that the Heisenberg picture signal of Eq. (\ref{Eq:out_signal_Heisenberg}) is equal to the interaction picture signal of Eq. (\ref{Eq:out_signal_interaction}), it suffices to show 
\begin{equation}
    \text{Tr}\big(\rho(-\infty) a^\dagger_{\text{pr, out}}(t_2)a_{\text{pr, out}}(t_1)\big) = \text{Tr}\big(\rho(\infty) a^\dagger_{\text{pr}}(t_2)a_{\text{pr}}(t_1)\big).
\label{Eq:A1}
\end{equation}
The presence of $\rho(\infty)$ in the interaction picture is not very intuitive, but this can be understood if we consider the following relation \cite{Ko_2022}:
\begin{equation}
    a_{\text{pr,out}}(t')=U^{\dagger}(t)a_{\text{pr}}(t')U(t),
\label{Eq:A2}
\end{equation}
where $t>t'$. $U(t)$ is defined below Eq. (\ref{Eq:Input_output_relation}). Taking a common time variable $t$, such that $t>t_1,t_2$. The left-hand side of Eq. (\ref{Eq:A1}) becomes
\begin{equation}
    \text{Tr}\big( \rho(-\infty)U^\dagger(t)a^\dagger_{\text{pr}}(t_2)U(t)U^\dagger(t)a_{\text{pr}}(t_1)U(t) \big).
\label{Eq:A3}
\end{equation}
Using the invariance of the trace under cyclic permutation and the unitary property $U(t)U^\dagger(t)=1$, (\ref{Eq:A3}) becomes
\begin{equation}
    \text{Tr}\big(U(t) \rho(-\infty)U^\dagger(t)a^\dagger_{\text{pr}}(t_2)a_{\text{pr}}(t_1) \big).
\end{equation}
Finally, taking $t\rightarrow \infty$, so that $U(t)\rho(-\infty)U^\dagger(t)\rightarrow \rho(\infty)$, we obtain the right-hand side of Eq. 
(\ref{Eq:A1}).

\section{Numerical parameters of Sec. \ref{sec:numerical_example}}
\label{app:numerical_parameter}
Using the two-state jump model in \cite{Schlawin_2016} for the matter system and following the notation in that work, we take $\omega_{fe}=11000\,\text{cm}^{-1}$, $\delta=200\,\text{cm}^{-1}$, $k=120\,\text{cm}^{-1}$, and $\gamma=100\,\text{cm}^{-1}$. From Eq. (19) of \cite{Schlawin_2016}, we derive the matter correlation function 
\begin{align}
\begin{split}
    \Tilde{F}(\omega',\omega;t_0) = e^{-i(\omega-\omega')t_0} \Big(&\frac{1}{(\omega-\omega'+i\gamma)}\frac{1}{(\omega-\omega_+ +2i\gamma)}\\
    &+\frac{2i\delta}{k+2i\delta}\frac{1}{(\omega-\omega'+i(k+\gamma))}\frac{1}{(\omega-\omega_- + i(k+2\gamma))}\\
    &-\frac{2i\delta}{k+2i\delta}\frac{1}{(\omega-\omega'+i(k+\gamma))}\frac{1}{(\omega-\omega_+ + 2i\gamma)}\Big),
\end{split}
\end{align}
where $\omega_\pm=\omega_{fe}\pm\delta$. We note that this is slightly different from Eq. (20) of \cite{Schlawin_2016}. We then multiply $\Tilde{F}(\omega',\omega;t_0)$ by a factor of $20$, so that around $10\%$ of the probe is absorbed at the peak of the pump-probe spectrum. The factor of $20$ effectively takes into account the light beam geometry, the molecular dipole strength, and the number of molecules in the sample.
\par The biphoton wavefunction of \cite{Schlawin_2016} takes the Gaussian form
\begin{equation}
    \Phi(\omega,\omega_r)=\mathcal{N}e^{-\frac{(\omega+\omega_r-\omega_0)^2}{2\sigma^2}}e^{-\beta[(\omega-\omega_0/2)T_2+(\omega_r-\omega_0/2)T_1]^2},
\end{equation}
where $\mathcal{N}$ is a normalization factor ensuring $\int d\omega d\omega_r |\Phi(\omega,\omega_r)|^2=1$, $\beta=0.04822$, $\omega_0=22000\,\text{cm}^{-1}$, $\sigma=1000\,\text{cm}^{-1}$, $T_1=-19.69\,\text{fs}$, and $T_2=70.31\,\text{fs}$. 
If we choose a fixed value of $\omega_r$, then the bivariate Gaussian biphoton wavefunction reduces to a single-variable Gaussian function $\propto e^{-(\omega-\omega'_0)^2/2\sigma'^2}$ with a modified center frequency 
\begin{equation}
    \omega'_0 = \big(\frac{1}{\sigma^2}+2\gamma T_2^2\big)^{-1}\bigg[\frac{\omega_0-\omega_r}{\sigma^2}+2\gamma T_2\big(\frac{\omega_0}{2}(T_1+T_2)-\omega_r T_1\big)\bigg]
\end{equation}
and variance 
\begin{equation}
    \sigma' = \big(\frac{1}{\sigma^2}+2\gamma T_2^2\big)^{-1/2} .
\end{equation}
This gives the explicit form of the quantum-inspired classical probe pulse corresponding to the biphoton pulse, which is thus seen to depend on the biphoton parameters 
$\gamma$, $T_1$, $T_2$, $\omega_0$ and $\sigma$, in addition to $\omega_r$. When $\omega_r=10400\,\text{cm}^{-1}$, the quantum-inspired pulse has  $\omega'_0=10874.81\,\text{cm}^{-1}$ and $\sigma'=236.09\,\text{cm}^{-1}$. When $\omega_r=11400\,\text{cm}^{-1}$, the quantum-inspired pulse has $\omega'_0=11083.46\,\text{cm}^{-1}$ and $\sigma'=236.09\,\text{cm}^{-1}$.

\newpage
\bibliographystyle{unsrt}
\typeout{}
\bibliography{references.bib}

\end{document}